\documentclass[10pt, draftclsnofoot, onecolumn]{IEEEtran}

\usepackage{setspace}

\usepackage{amsfonts}
\usepackage{color}
\usepackage{amsmath}
\usepackage{multirow}
\usepackage{graphicx}
\usepackage{amsfonts}
\usepackage{amsmath,epsfig}
\usepackage{stfloats}
\usepackage{amssymb}
\usepackage{url}
\usepackage{bm}
\usepackage{cite}
\usepackage{amsmath}
\usepackage{amssymb}
\usepackage{latexsym}
\usepackage{multirow}
\usepackage{epsfig}
\usepackage{graphics}
\usepackage{mathrsfs}
\usepackage{algorithmic}
\usepackage{algorithm}
\usepackage[all]{xy}

\usepackage{pifont}
\usepackage{bbding}
\usepackage{stmaryrd}
\usepackage{amssymb}
\usepackage{amsfonts}
\usepackage{epic}
\usepackage{stfloats}
\usepackage{epstopdf}
\usepackage{bm}
\usepackage{xcolor}
\usepackage{mathrsfs}
\usepackage{pifont}
\usepackage{bbding}
\usepackage{amsmath,epsfig}
\usepackage{stmaryrd}
\usepackage{amsmath}
\usepackage{amssymb}
\usepackage{amsthm}

\usepackage{amsfonts}
\usepackage{epic}
\usepackage{graphicx}
\usepackage{enumerate}
\usepackage{latexsym}
\usepackage{epstopdf}
\usepackage{multirow}
\usepackage{stfloats}
\usepackage{bm}
\usepackage{booktabs}
\usepackage{color}
\usepackage{cases}
\usepackage{array}
\usepackage{makecell}
\usepackage{times}

\usepackage{diagbox}
\usepackage{slashbox}
\usepackage{booktabs}

\ifCLASSOPTIONcompsoc
\usepackage[caption=false,font=normalsize,labelfont=sf,textfont=sf]{subfig}
\else
\usepackage[caption=false,font=footnotesize]{subfig}
\fi

\allowdisplaybreaks

\theoremstyle{definition} 

\begin{document}

\title{Throughput Maximization for IRS-Assisted Wireless Powered Hybrid NOMA and TDMA}

\author{Dingcai~Zhang,~Qingqing~Wu,~\IEEEmembership{Member,~IEEE},~Miao~Cui,~Guangchi~Zhang,~and~Dusit~Niyato,~\IEEEmembership{Fellow,~IEEE}
\thanks{D. Zhang, M. Cui, and G. Zhang are with the School of Information Engineering, Guangdong University of Technology, Guangzhou, China (email: dingcaizhang@126.com, cuimiao@gdut.edu.cn, gczhang@gdut.edu.cn). Q. Wu is with the State Key Laboratory of Internet of Things for Smart City, University of Macau, Macau, China (email: qingqingwu@um.edu.mo). D. Niyato is with the School of Computer Science and Engineering, Nanyang Technological University, Singapore 639798 (email: dniyato@ntu.edu.sg). G. Zhang is the corresponding author. }}

\maketitle

\begin{abstract}
The high reflect beamforming gain of the intelligent reflecting surface (IRS) makes it appealing not only for wireless information transmission but also for wireless power transfer. In this letter, we consider an IRS-assisted wireless powered communication network, where a base station (BS) transmits energy to multiple users grouped into multiple clusters in the downlink, and the clustered users transmit information to the BS in the manner of hybrid non-orthogonal multiple access and time division multiple access in the uplink. We investigate optimizing the reflect beamforming of the IRS and the time allocation among the BS's power transfer and different user clusters' information transmission to maximize the throughput of the network, and we propose an efficient algorithm based on the block coordinate ascent, semidefinite relaxation, and sequential rank-one constraint relaxation techniques to solve the resultant problem. Simulation results have verified the effectiveness of the proposed algorithm and have shown the impact of user clustering setup on the throughput performance of the network.
\end{abstract}


\section{Introduction}
Intelligent reflecting surface (IRS) can adaptively tune the phase shifts of its reflected signals by a large number of low-cost reflecting elements integrated on it to achieve high reflect beamforming gain \cite{Wu2021, Renzo2019}. Therefore, leveraging IRS's intelligent reflection has been regarded as a promising way to improve the spectrum and energy efficiency of future wireless communication networks \cite{Wu2019, Pan2020TWC}. On the other hand, wireless power transfer has been introduced into the rapidly developing Internet-of-Things (IoT) networks to resolve the network devices' energy limitation problem. Such networks are also called wireless power communication networks (WPCNs) \cite{Wu2018}, which provide an efficient way to realize sustainable IoT networks. Recently, both the transmit power minimization study in \cite{WuJSAC2020} and the weighted sum-rate maximization study in \cite{Pan2020} have unveiled that IRSs can achieve high wireless power transfer and information transmission efficiency due to the high reflective beamforming gain. Therefore, integrating IRSs with WPCN is a concrete step toward the realization of low-cost sustainable IoT networks. In \cite{Zheng2020}, the throughput maximization of a two-user IRS-assisted WPCN has been investigated, where IRS reflect beamforming, power allocation and time allocation have been jointly optimized. In \cite{Lyu2020}, joint optimization of IRS reflect beamforming and network resource allocation for throughput maximization have been investigated in a multiuser IRS-assisted WPCN.

Meanwhile, non-orthogonal multiple access (NOMA) has been proposed to enable the access of massive numbers of devices in IoT networks \cite{Ding2014}. A recent study in \cite{Zeng2020} shows that IRS-assisted NOMA can achieve a higher uplink sum rate than that of the IRS-assisted orthogonal multiple access (OMA). However, the receiver complexity required by using successive interference cancellation (SIC) in NOMA is significantly higher than that of OMA. Furthermore, the error propagation incurred by subtracting the signals of previous users during the process of SIC will constitute a performance-limiting factor when the number of users is large. Furthermore, NOMA may not always outperform OMA (e.g. time division multiple access (TDMA)) in the IRS-assisted communication systems when the IRS can be dynamically switched \cite{Zhu2020}, and OMA may even have higher energy efficiency than that of the NOMA in WPCN \cite{Wu2018}. Therefore, how to achieve a trade-off between complexity and performance in multiuser IRS-assisted WPCNs is still a challenge.

\begin{figure}[!t]
	\centering
	\includegraphics[width=0.5\textwidth]{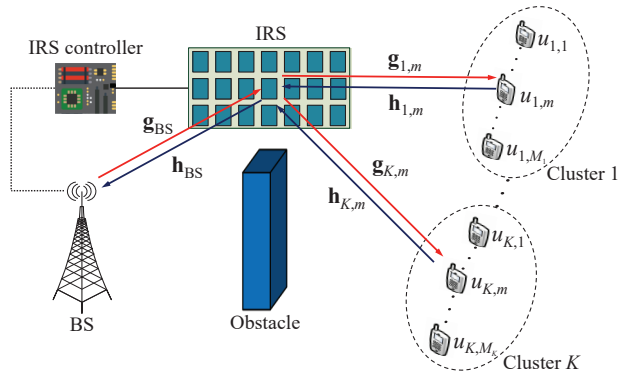}
	\caption{An IRS-assisted wireless powered hybrid NOMA and TDMA network.} 
	\label{FigSystem}
\end{figure}

In this letter, we consider an IRS-assisted WPCN, as shown in Fig. \ref{FigSystem}, where multiple users harvest energy and transmit information from/to a base station (BS) with the assistance of an IRS. Different from existing works on IRS-assisted WPCNs, a novel {\it hybrid} NOMA and TDMA scheme is proposed to balance complexity and performance. In particular, for the uplink information transmission, the users are grouped into several clusters such that different clusters of users transmit information at different times using TDMA, while the users in the same cluster transmit information simultaneously using NOMA. The proposed scheme can be regarded as a generalization of NOMA and TDMA, which not only decreases the complexity of implementing SIC but also offers additional degrees of freedom to improve performance. We aim to maximize the throughput of the users' information transmission by optimizing the time allocation among the BS's energy transfer and different user clusters' information transmission, as well as the IRS reflect beamforming during the BS and the users' transmission periods. Although the considered optimization problem is non-convex and challenging, we propose an efficient algorithm to solve it suboptimally by applying the block coordinate ascent (BCA), semidefinite relaxation (SDR), and sequential rank-one constraint relaxation (SROCR) techniques. Simulation results have verified the performance superiority of the proposed algorithm, as compared to some benchmark algorithms, and have shown the impact of user grouping and clustering setup on the throughput performance.


\section{System Model and Problem Formulation}
As shown in Fig. \ref{FigSystem}, we consider an IRS-assisted WPCN, which includes a single-antenna BS, $M$ single-antenna users, and an IRS with $N$ reflecting elements. The direct links between the BS and the users are blocked by obstacles, so the IRS is deployed to assist the power transfer and information transmission of the network. Note that the proposed algorithm can be applied directly to the case with direct links or with a multi-antenna BS. To provide a performance upper bound benchmark, we assume that the channel state information (CSI) of all links is perfectly estimated and known by the BS \cite{Wu2021}, and the trade-off between channel estimation and performance optimization will be left for future work.

In the downlink power transfer (DPT) phase, the BS broadcasts energy signal to the users for a duration $\tau_0$, and the IRS reflects the energy signal with a reflect beamforming matrix $\Phi_0 \triangleq \text{diag}(\mathbf{w}_0)$, where $\mathbf{w}_0 \triangleq [w_{1,0}, \ldots, w_{N,0}]^T$, and the operator $\text{diag}(\mathbf{a})$ forms a diagonal matrix with the elements of a vector $\mathbf{a}$. For $n=1,\ldots,N$, $w_{n,0}=\alpha_{n,0}e^{j\theta_{n,0}}$, and $\alpha_{n,0}\in [0,1]$ and $\theta_{n,0}\in [0,2\pi)$ are the amplitude and phase shift coefficient of the IRS's $n$th reflecting element during the DPT phase, respectively\footnote{We assume that the phase shifts are continuous, and the obtained performance can be used as a benchmark for the performance of the systems with discrete phase shifts. The impact of discrete phase shifts are evaluated in Section IV.}. To achieve the maximum reflecting power gain of the IRS, we assume the reflection amplitudes of all elements are one, i.e., $\alpha_{n,0} = |w_{n,0}| = 1$, $\forall n$ \cite{Wu2019, Cui2019}.

In the uplink information transmission (UIT) phase, the users transmit information to the BS with a hybrid NOMA and TDMA scheme, where all $M$ users are grouped into $K$ clusters with $M_k$ users in the $k$th cluster such that $\sum_{k=1}^K M_k = M$. Different clusters of users transmit information at different times in the order of their cluster index, while users in the same cluster transmit simultaneously. During the transmission of the $k$th cluster of users, whose duration is $\tau_k$, the IRS reflects the user signal with a reflect beamforming matrix $\Phi_k \triangleq \text{diag}(\mathbf{w}_k)$, where $\mathbf{w}_k \triangleq [w_{1,k}, \ldots, w_{N,k}]^T$, and similar to $\mathbf{w}_0$, $|w_{n,k}| = 1$, $\forall n,k$, is satisfied. Note that the hybrid NOMA and TDMA scheme becomes TDMA when $K=M$ and $M_k=1, \forall k$, and becomes NOMA when $K=1$ and $M_1=M$ \cite{Wu2018}.

We consider a block quasi-static flat-fading channel model by assuming that all channels are constant over each power transfer and information transmission block and may vary over different blocks. The total duration of one block is denoted by $T$. Let $u_{k,m}$ denote the $m$th user in the $k$th cluster, and the channel coefficients from the BS to the IRS, from the IRS to $u_{k,m}$, from $u_{k,m}$ to the IRS, from the IRS to the BS are denoted by $\mathbf{g}_{\text{BS}} \in \mathbb{C}^{N \times 1}$, $\mathbf{g}_{{k,m}} \in \mathbb{C}^{N \times 1}$, $\mathbf{h}_{{k,m}} \in \mathbb{C}^{N \times 1}$, $\mathbf{h}_{\text{BS}} \in \mathbb{C}^{N \times 1}$, respectively, where $\mathbb{C}^{x \times y}$ denotes the set of $x \times y$ complex-valued matrices. Let $P_0$ denote the transmit power of the BS, and then the harvested energy of $u_{k,m}$ can be expressed by
\begin{equation} \label{EquHarEnerg}
E_{k,m}= \eta \tau_0 P_0 |\mathbf{g}^{H}_{{k,m}} \Phi_0 \mathbf{g}_{\text{BS}}|^2,
\end{equation}
where $\eta$ denotes the energy harvesting efficiency of the users\footnote{In \eqref{EquHarEnerg}, we adopt a linear energy harvesting model that is commonly used in existing works \cite{WuJSAC2020, Pan2020, Zheng2020} by considering the harvested energy is in the linear regime of the energy harvester. How to extend our work to the scenario of non-linear energy harvester is left for future work.}.  With the harvested energy, the transmit power of $u_{k,m}$ is $P_{k,m} = E_{k,m}/\tau_k$. Thus, the total throughput of the $k$th user cluster in bits/Hertz (bits/Hz) can be written as \cite{Wu2018}
\begin{equation} \label{Equ_throu_cluster}
R_k = \tau_k \log_2 \Bigg( 1 + \frac{\sum\limits_{m=1}^{M_k} P_{k,m}|\mathbf{h}^{H}_{\text{BS}} \Phi_k \mathbf{h}_{k,m}|^2 }{\sigma^2} \Bigg),
\end{equation}
where $\sigma^2$ denotes the noise power at the BS. Accordingly, the throughput of all users' information transmission in bits/Hz is $R = \sum_{k=1}^K R_k $. It is observed from \eqref{EquHarEnerg}--\eqref{Equ_throu_cluster} that the durations of power transfer and information transmission and the IRS reflect beamforming have a great impact on the value of $R$. 

In this letter, we aim to maximize the throughput $R$ by jointly optimizing the time allocation among the BS's energy transfer and different user clusters' information transmission, i.e., $\tau_0$ and $\{\tau_k\}$, as well as the IRS reflect beamforming at different time, i.e., $\mathbf{w}_0$ and $\{\mathbf{w}_k\}$. The considered problem can be formulated as
\begin{subequations}  \label{EquOriProb}
	\begin{align}
		\text{(P1)}:  \max \limits_{\tau_0,\{\tau_k\},\mathbf{w}_0,\{ \mathbf{w}_k\}} & \sum_{k=1}^K  R_k \label{EquObjFun}   \\
	\textrm{s.t.}\; \quad \quad & |w_{n,0}| =1, \; \forall n,   \label{EquIRSCon0}  \\
	&|w_{n,k}| =1, \; \forall n, k, \label{EquIRSConk}  \\
	&\tau_0 + \sum\limits_{k=1}^K \tau_k \leq T, \label{EquTotalDurCon} \\
	& \tau_0 \geq 0,  \; \tau_k \geq 0, \; \forall k.  \label{EquDurPos}
	\end{align}
\end{subequations}
In problem (P1), the optimization variables $\tau_0$, $\{\tau_k\}$, $\mathbf{w}_{0}$, and $\{\mathbf{w}_{\text{k}}\}$ are intricately coupled in the objective function, and the objective function is not jointly concave with respect to them. Furthermore, the left-hand-sides (LHSs) of the equality constraints \eqref{EquIRSCon0} and \eqref{EquIRSConk} are non-linear functions, so (P1) is a highly non-convex optimization problem whose optimal solution is difficult to obtain in polynomial time. Nevertheless, we propose an efficient algorithm to obtain its suboptimal solution in the next section.

\section{Proposed Algorithm}
It is observed that in (P1), \eqref{EquIRSCon0} is constraints on $\mathbf{w}_{0}$, \eqref{EquIRSConk} is constraints on $\{\mathbf{w}_{\text{k}}\}$, and \eqref{EquTotalDurCon} and \eqref{EquDurPos} are constraints on $\tau_0$ and $\{\tau_k\}$. This motivates us to resolve the optimization variable coupling in the objective function of (P1) by using the BCA technique, which divides the optimization variables of (P1) into three blocks, i.e., $\mathbf{w}_{0}$, $\{\mathbf{w}_{k}\}$, and $( \tau_0, \{\tau_k\})$, and alternately optimizes one block with the others fixed until achieving convergence. Specifically, the following three sub-problems need to be solved: sub-problem 1 optimizes $\mathbf{w}_{0}$ under given $\{\mathbf{w}_{k}\}$, $\tau_0$ and $\{\tau_k\}$, sub-problem 2 optimizes $\{\mathbf{w}_{k}\}$ under given $\mathbf{w}_{0}$, $\tau_0$ and $\{\tau_k\}$, and sub-problem 3 optimizes $\tau_0$ and $\{\tau_k\}$ under given $\mathbf{w}_{0}$ and $\{\mathbf{w}_{k}\}$. The procedures of solving all sub-problems and the overall proposed algorithm are presented below.

\subsection{Optimizing $\mathbf{w}_{0}$ with Given $\{\mathbf{w}_{\text{k}}\}$, $\tau_0$ and $\{\tau_k\}$}
By letting $\lambda_k=\frac{\eta\tau_0P_0}{\tau_k\sigma^2}$, $b_{k,m} = |\mathbf{h}^{H}_{\text{BS}} \Phi_k \mathbf{h}_{k,m}|^2$, $\mathbf{\hat{g}}_{k,m} = \mathbf{g}_{k,m}\odot\mathbf{g}_{\text{BS}}$, where $\odot$ denotes the Hadamard product operator, sub-problem 1 can be formulated as
\begin{subequations}
\begin{align}
\text{(P2)}:  \max \limits_{\mathbf{w}_0}& \sum_{k=1}^K\tau_k\log_2 \bigg( 1 + \lambda_k \sum_{m=1}^{M_{k}}|\mathbf{w}_{0}^{H} \mathbf{\hat{g}}_{k,m}|^2b_{k,m}\bigg) \label{EquOriProbObj} \\
\textrm{s.t.}\;\;&|w_{n,0}| = 1, \; \forall n.  \label{EquConwInq2}
\end{align}
\end{subequations}
Problem (P2) is difficult to solve since its objective function is non-concave with respect to $\mathbf{w}_0$, and the LHSs of \eqref{EquConwInq2} is non-linear. We apply the SDR technique to overcome such a difficulty. First, we write the summation term in the logarithmic function of \eqref{EquOriProbObj} into the following form:
\begin{equation}
\sum_{m=1}^{M_{k}}|\mathbf{w}_{0}^{H} \mathbf{\hat{g}}_{k,m}|^2b_{k,m}=\mathbf{w}_{0}^{H} \mathbf{G}_{k}\mathbf{w}_{0} = \text{tr}(\mathbf{G}_{k}\mathbf{w}_{0}\mathbf{w}_{0}^{H} ),
\end{equation}
where $\mathbf{G}_{k}=\sum_{m=1}^{M_{k}}b_{k,m}\mathbf{\hat{g}}_{k,m}\mathbf{\hat{g}}^{H}_{k,m}$, and $\text{tr}(\cdot)$ denotes the trace operator. Besides, we write constraint \eqref{EquConwInq2} into
\begin{equation}
	\mathbf{w}_{0}^{H} \mathbf{B}_n \mathbf{w}_{0} = \text{tr}(\mathbf{B}_{n}\mathbf{w}_{0}\mathbf{w}_{0}^{H} ) = 1, \; \forall n,
\end{equation}
where the matrix $\mathbf{B}_n$ is constructed by letting its $(i,j)$th element, denoted by $[\mathbf{B}_n]_{i,j}$, satisfy
\begin{equation}
[\mathbf{B}_n]_{i,j} = \begin{cases}
1 & i=j=n  \\  0 & \text{otherwise}.
\end{cases}
\end{equation}

Then, we let $\mathbf{W}_{0} \triangleq \mathbf{w}_{0} \mathbf{w}_{0}^{H}$ and re-express problem (P2) as
\begin{subequations}
\begin{align}
\text{(P3)}:  \max \limits_{\mathbf{W}_0 \succeq 0 } &\sum_{k=1}^K\tau_k\log_2 \bigg( 1 + \lambda_k \text{tr}(\mathbf{G}_{k}\mathbf{W}_{0})\bigg)  \\
\textrm{s.t.}\;\;&\text{tr}(\mathbf{B}_{n}\mathbf{W}_{0})=1, \; \forall n    \label{EquConModuOne}   \\
&\text{rank}(\mathbf{W}_{0}) =1. \label{Rank2_one}
\end{align}
\end{subequations}
Although problem (P3) is NP-hard due to the rank-one constraint \eqref{Rank2_one}, we propose an SROCR-based algorithm to solve it. Unlike the existing approach for the (P3)-type problems \cite{Wu2019, Cui2019}, which first solves its relaxed problem that ignores the rank-one constraint and then constructs a feasible solution based on the solution to the relaxed problem by using e.g., Gaussian randomization method, the SROCR-based algorithm relaxes the rank-one constraint gradually over iterations and can find a locally optimal solution to (P3) eventually \cite{Cao2017}. Suppose that $\mathbf{W}_{0}^{(i)}$ is the obtained solution in iteration $i$. In iteration $i+1$, the proposed algorithm constructs and solves the following relaxed problem:
\begin{subequations}
	\begin{align}
	\text{(P4)}: & \max \limits_{\mathbf{W}_0 \succeq 0 } \sum_{k=1}^K\tau_k\log_2 \bigg( 1 + \lambda_k \text{tr}(\mathbf{G}_{k}\mathbf{W}_{0})\bigg)  \\
	\textrm{s.t.}\;& \eqref{EquConModuOne},   \\
&  \mathbf{u}_{\max}(\mathbf{W}_{0}^{(i)})^H \mathbf{W}_{0} \mathbf{u}_{\max}(\mathbf{W}_{0}^{(i)}) \geq v^{(i)}\text{tr}(\mathbf{W}_{0}),  \label{SROCR}  \end{align}
\end{subequations}
where $\mathbf{u}_{\max}(\mathbf{W}_{0}^{(i)})$ denotes the eigenvector corresponding to the largest eigenvalue of $\mathbf{W}_{0}^{(i)}$, and $v^{(i)}$ denotes an introduced relaxation parameter. Problem (P4) is a semidefinite programming (SDP) problem and can be efficiently solved by the interior-point method. The relaxation from \eqref{Rank2_one} to \eqref{SROCR} becomes tighter and tighter, as $v^{(i)}$ increases from $0$ to $1$ over iterations. When $v^{(i)}=1$, \eqref{SROCR} is equivalent to \eqref{Rank2_one}, and thus the solution obtained by solving (P4) is a solution to (P3). The detail of the proposed algorithm is presented in Algorithm 1, where $\lambda_{\max}(\mathbf{W}_{0})$ denotes the largest eigenvalue of $\mathbf{W}_{0}$, $g_0(\mathbf{W}_{0})$ denotes the objective value of (P4) with solution $\mathbf{W}_{0}$, and $\epsilon_1$ and $\epsilon_2$ are thresholds indicating the convergence accuracy. After obtaining the solution to (P3), denoted by $\mathbf{W}_{0}^*$, the solution to sub-problem 1 can be obtained by a rank-one decomposition on $\mathbf{W}_{0}^*$.

\begin{algorithm}[!t]
	\caption{SROCR-Based Algorithm for Problem (P3).}
	\begin{algorithmic}[1]
		\STATE \textbf{Initialization:} Solve problem (P4) when $v^{(0)}=0$ and denote the solution by $\mathbf{W}_{0}^{(0)}$. Define an initial step size $\delta^{(0)} \in \big(0, 1 - \frac{ \lambda_{\max}(\mathbf{W}_{0}^{(0)}) }{\text{tr}(\mathbf{W}_{0}^{(0)})} \big]$ and set $i=0$.
		\REPEAT
		\STATE Solve problem (P4) under given $\{v^{(i)},\mathbf{W}_{0}^{(i)}\}$.
		\IF{Problem (P4) is solvable}
		\STATE Denote the solution by $\mathbf{W}_{0}^{(i+1)}$ and set $\delta^{(i+1)}=\delta^{(i)}$.
		\ELSE
		\STATE Set $\mathbf{W}_{0}^{(i+1)}=\mathbf{W}_{0}^{(i)}$ and $\delta^{(i+1)}=\delta^{(i)}/2$.
		\ENDIF
		\STATE Set $v^{(i+1)} = \min \big(1, \frac{ \lambda_{\max}(\mathbf{W}_{0}^{(i+1)}) }{\text{tr}(\mathbf{W}_{0}^{(i+1)})} +\delta^{(i+1)}\big)$.
		\STATE Set $i=i+1$.		
		\UNTIL $v^{(i-1)} \geq \epsilon_1$ \& $|g_0(\mathbf{W}_{0}^{(i)}) - g_0(\mathbf{W}_{0}^{(i-1)})| \leq \epsilon_2$.
	\end{algorithmic}
\end{algorithm}

\subsection{Optimizing $\{\mathbf{w}_{\text{k}}\}$ with Given $\mathbf{w}_{0}$, $\tau_0$ and $\{\tau_k\}$ }
By letting $c_{k,m} = |\mathbf{w}_{0}^{H} \mathbf{\hat{g}}_{k,m}|^2$ and $\mathbf{\hat{h}}_{k,m} = \mathbf{h}_{\text{BS}}\odot\mathbf{h}_{k,m}$, sub-problem 2 can be formulated as
\begin{subequations}
\begin{align}
\text{(P5)}:  \max \limits_{\{ \mathbf{w}_k \} }& \sum_{k=1}^K\tau_k\log_2 \bigg( 1 + \lambda_k \sum_{m=1}^{M_{k}}|\mathbf{w}^{H}_{k}\mathbf{\hat{h}}_{k,m}|^2c_{k,m}\bigg)  \\
\textrm{s.t.}\;\;&|w_{n,k}| = 1, \; \forall n,k.
\end{align}
\end{subequations}
It is worth noting that the objective function and constraints of problem (P5) can be decoupled with respect to different user clusters, which means that it can be solved by optimizing $\mathbf{w}_1, \ldots, \mathbf{w}_K$ independently. In particular, let $\mathbf{H}_{k}=\sum_{m=1}^{M_{k}}c_{k,m}\mathbf{\hat{h}}_{k,m}\mathbf{\hat{h}}^{H}_{k,m}$ and $\mathbf{W}_{k} \triangleq \mathbf{w}_{k}\mathbf{w}^{H}_{k}$ and the problem optimizing $\mathbf{w}_k$ can be reduced to
\begin{subequations}
\begin{align}
\text{(P6)}:  \max \limits_{ \mathbf{W}_k \succeq 0 } \; \;&\text{tr}(\mathbf{H}_{k}\mathbf{W}_{k}) \\ 
\textrm{s.t.}\;\;&\text{tr}(\mathbf{B}_{n}\mathbf{W}_{k})=1, \; \forall n,     \\
&\text{rank}(\mathbf{W}_{k}) =1 \label{EquRankWk}.
\end{align}
\end{subequations}
Problem (P6) can be solved similarly by an SROCR-based algorithm like Algorithm 1, and the detail is omitted here for brevity.

\subsection{Optimizing $\tau_0$ and $\{\tau_k\}$ with Given $\mathbf{w}_{0}$ and $\{\mathbf{w}_{\text{k}}\}$}
Sub-problem 3 can be formulated as
\begin{subequations}
\begin{align}
\text{(P7)} : \max \limits_{\tau_0,\{\tau_k\}} & \sum\limits_{k=1}^K\tau_k\log_2  \Bigg( 1 + \frac{  \eta\tau_0 P_0\sum\limits_{m=1}^{M_{k}} b_{k,m} c_{k,m}}{ \tau_k \sigma^2} \Bigg)  \\
\text{s.t.}\;\;& \eqref{EquTotalDurCon}, \eqref{EquDurPos}.
\end{align}
\end{subequations}
Problem (P7) is a convex optimization problem, so by analyzing its Karush-Kuhn-Tucker (KKT) conditions, its optimal solution can be obtained as
\begin{equation}   \label{EquTauSol}
\tau_0^*= \frac{T}{1+\sum\nolimits_{k=1}^K\frac{\gamma_k}{ x_k^*}}, \; \; \tau_k^*=\frac{\gamma_k}{ x_k^*}\tau_0^*, \; \forall k,
\end{equation}
where $\gamma_k= \frac{ \eta P_0 }{ \sigma^2 } \sum\nolimits_{m=1}^{M_{k}} b_{k,m} c_{k,m} $ and $x_k^*$ is the solution of the following equation
\begin{equation}  \label{EquEqux}
\log_2(1+x_k)-\frac{x_k\log_2(e)}{1+x_k} -\sum\limits_{k=1}^K\frac{\gamma_k \log_2(e)}{1+x_k} = 0.
\end{equation}
Note that $x_k^*$ can be obtained by a numerical method such as the bisection method. The procedure of obtaining \eqref{EquTauSol} is similar to that in \cite{Wu2018} and thus omitted here for simplicity.

\begin{algorithm}[!t]
	\caption{Proposed IRS Reflect Beamforming and Time Allocation Algorithm.}
	\begin{algorithmic}[1]
		\STATE \textbf{Initialization:} Set initial value for $\mathbf{w}_{0}$, $\{ \mathbf{w}_k \}$, $\tau_0$, and $\{ \tau_{k} \}$. Set $l=0$ and $R^{(0)} = f ( \mathbf{w}_{0}, \{ \mathbf{w}_{k} \}, \tau_0, \{ \tau_{k} \} )$.
		\REPEAT
		\STATE Set $l=l+1$.
		\STATE Under given $\{\mathbf{w}_k\}$, $\tau_0$ and $\{ \tau_{k} \}$, update $\mathbf{w}_{0}$ by solving problem (P3).
		\STATE Under given $\mathbf{w}_0$, $\tau_0$ and $\{ \tau_{k} \}$, update $\{\mathbf{w}_k\}$ by solving problem (P6).
	    \STATE Under given $\mathbf{w}_{0}$ and $\{ \mathbf{w}_k \}$, update $\tau_0$ and $\{ \tau_{k} \}$ by solving problem (P7).
        \STATE Set $R^{(l)} = f ( \mathbf{w}_{0}, \{ \mathbf{w}_{k} \}, \tau_0, \{ \tau_{k} \} )$.
		\UNTIL {$\frac{R^{(l)} - R^{(l-1)}}{R^{(l)}} < \epsilon$.}
	\end{algorithmic} 
\end{algorithm}  

\subsection{Overall Algorithm and Complexity Analysis}
The overall algorithm for problem (P1) is present in Algorithm 2, where $f( \mathbf{w}_{0}, \{ \mathbf{w}_{k} \}, \tau_0, \{ \tau_{k} \} )$ denotes the objective value of (P1) with solution $\mathbf{w}_{0}$, $\{ \mathbf{w}_{k} \}$, $\tau_0$, and $\{ \tau_{k} \}$ and $\epsilon$ is a small positive threshold indicating the convergence accuracy. The algorithm solves sub-problems 1--3 in steps 4--6, respectively. Since the obtained solutions to sub-problems 1 and 2 are locally optimal and that to sub-problem 3 is globally optimal, the objective value of (P1) over iterations is non-decreasing. Furthermore, it must be upper-bounded by a finite value, so Algorithm 2 will converge. The main complexities are in executing steps 4 and 5, for which the complexities are $\mathcal{O}(N^{4.5})$ and $\mathcal{O}(KN^{4.5})$, respectively. Thus, the complexity of Algorithm 2 is $\mathcal{O}( (K+1) N^{4.5} )$. Note that Algorithm 2 can be executed at the BS, and the obtained result can be sent to the IRS and users over a reliable control link \cite{Wu2021}.

\section{Simulation Results}
This section provides numerical results to validate the effectiveness of the proposed algorithm. The simulation parameters are set as follows. There are $M=12$ users, and the total block length is set as $T=0.1$ s. The channel coefficients are set as $\mathbf{g}_{\text{BS}} =\sqrt{\zeta_0 ( d_0 / d_{\text{BI}} )^{ \alpha_{\text{BI}} } }\mathbf{\tilde{g}}_{\text{BS}}$, $\mathbf{g}_{k,m} =\sqrt{\zeta_0 ( d_0 / d_{k,m} )^{ \alpha_{k,m}}}\mathbf{\tilde{g}}_{k,m}$, $\mathbf{h}_{\text{BS}} =\sqrt{\zeta_0 ( d_0 / d_{\text{BI}} )^{ \alpha_{\text{BI}}}}\mathbf{\tilde{h}}_{\text{BS}}$, and $\mathbf{h}_{k,m} =\sqrt{\zeta_0 ( d_0 / d_{k,m} )^{ \alpha_{k,m}}}\mathbf{\tilde{h}}_{k,m}$, where $\zeta_0=-30$ dB denotes the path loss at the reference distance $d_0= 1$ m, $d_{\text{BI}}$ and $d_{k,m}$ denote the distance between the BS and the IRS and that between $u_{k,m}$ and the IRS, respectively, $\alpha_{\text{BI}}=2.2$ and $\alpha_{k,m}=2.5$ are the path loss exponents, and $\mathbf{\tilde{g}}_{\text{BS}}$, $\mathbf{\tilde{g}}_{k,m}$, $\mathbf{\tilde{h}}_{\text{BS}}$, $\mathbf{\tilde{h}}_{k,m}$ are small-scale fading components modeled by the Rician fading model with Rician factors setting as $1$ \cite{Cui2019}. The other parameters are set as $d_{\text{BI}}=1$ m, $\sigma^2= -110$ dBm, $\epsilon_1=0.95$, $\epsilon_2=10^{-3}$, and $\epsilon = 10^{-3}$.

We denote the proposed algorithm by `Optimized IRS w/ time allocation (TA)', and compare it to the following benchmark algorithms. 1) `Optimized IRS w/o TA': it fixes the durations as $\tau_0 = \tau_k = T/(K+1)$, $\forall k$, and optimizes IRS reflect beamforming by using steps 4-5 of Algorithm 2. 2) `Random IRS w/ TA': it uses random IRS reflect beamforming and optimizes $\tau_0$ and $\{\tau_k\}$ by using step 6 of Algorithm 2. 3) `Random IRS w/o TA': it uses random IRS reflect beamforming and sets $\tau_0 = \tau_k = T/(K+1)$, $\forall k$. 4) `Optimized IRS w/ TA (same IRS)': it uses the same IRS reflect beamforming over the whole block, i.e., $\mathbf{w}_0=\mathbf{w}_k$, $\forall k$, and optimizes IRS beamforming and time allocation by an iterative algorithm similar to Algorithm 2. 5) `Upper Bound': it constructs relaxed problems of (P3) and (P6) by dropping the rank-one constraints \eqref{Rank2_one} and \eqref{EquRankWk} from them, respectively, and solves the relaxed problems of (P3) and (P6), as well as (P7) alternatively to obtain a throughput upper bound of the proposed algorithm. 6) `Discrete phase': it is similar to the `Optimized IRS w/ TA' algorithm except that the phase shift of each IRS reflecting element is discretized into $2^b$ levels over the interval $[0, 2\pi)$ uniformly, where $b$ denotes the control bit number for each reflecting element. The following results are obtained by averaging the performances of 1000 random channel realizations.

\begin{figure*}[!t]
\centering
\subfloat[Throughputs versus $N$.]{\includegraphics[width=0.23\textwidth]{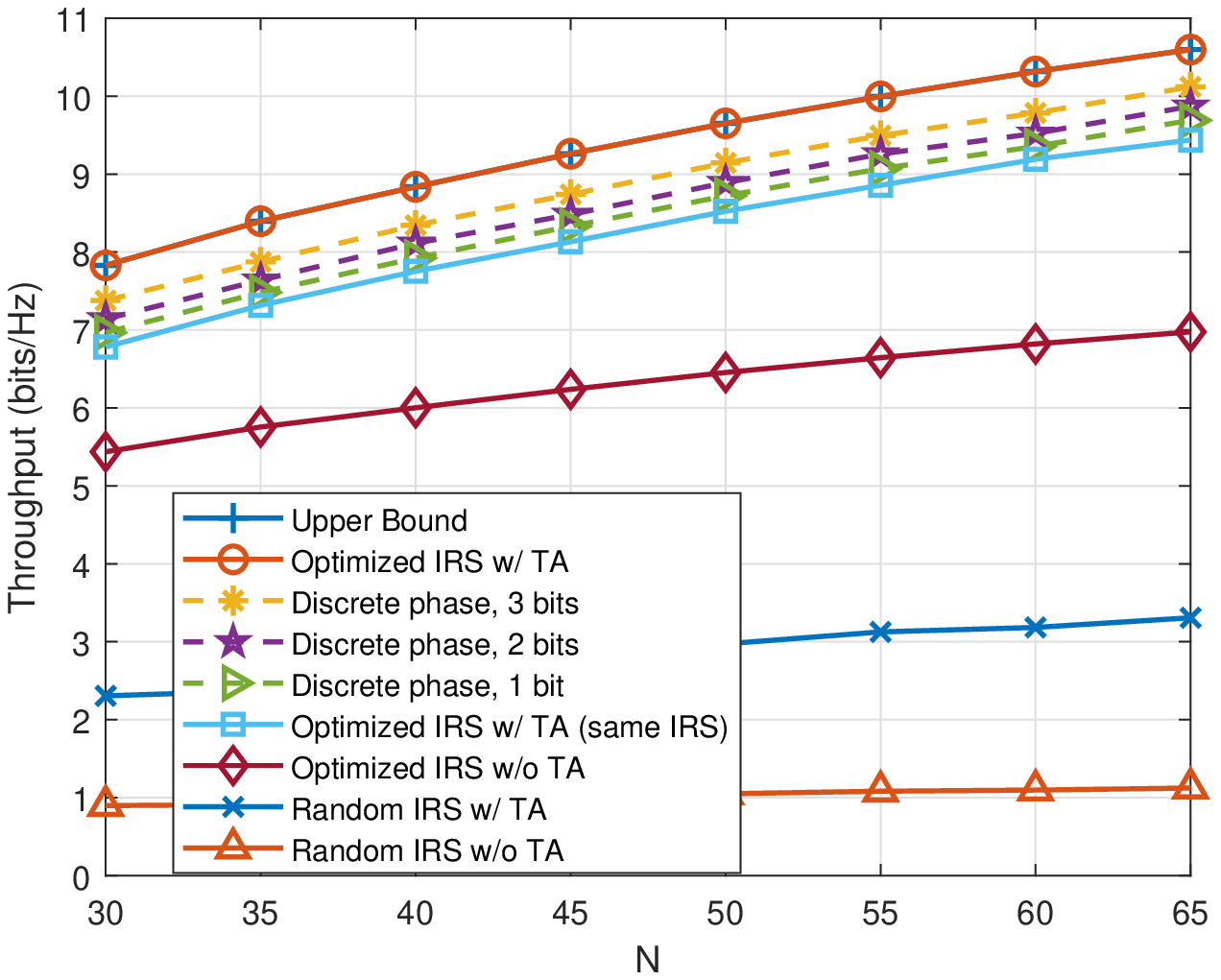}
\label{FigThr_N}} 
\hfil
\subfloat[Throughputs versus $d_{k,m}$.]{\includegraphics[width=0.23\textwidth]{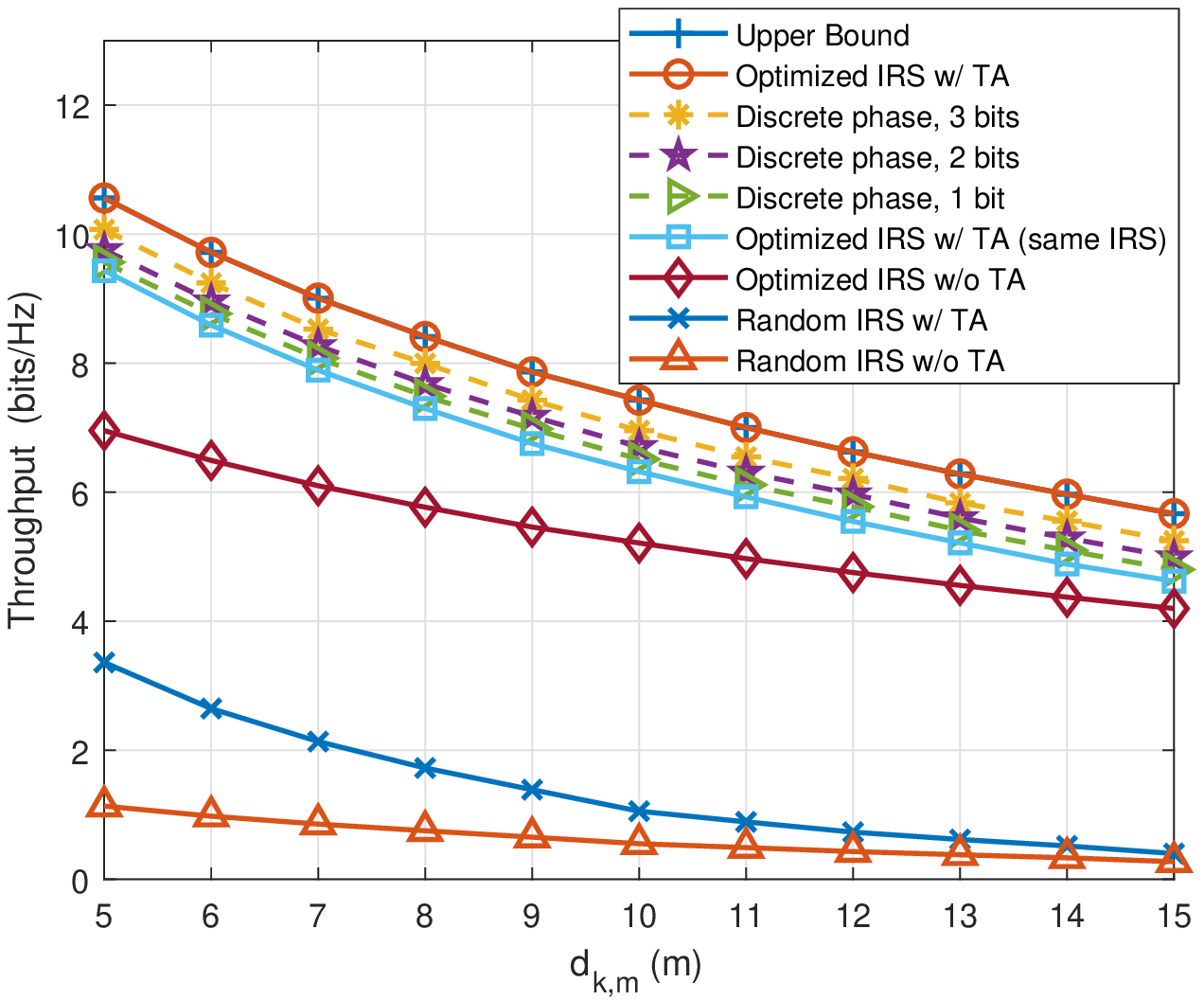}
\label{FigThr_D}} 
\hfil
\subfloat[Impact of user grouping.]{\includegraphics[width=0.23\textwidth]{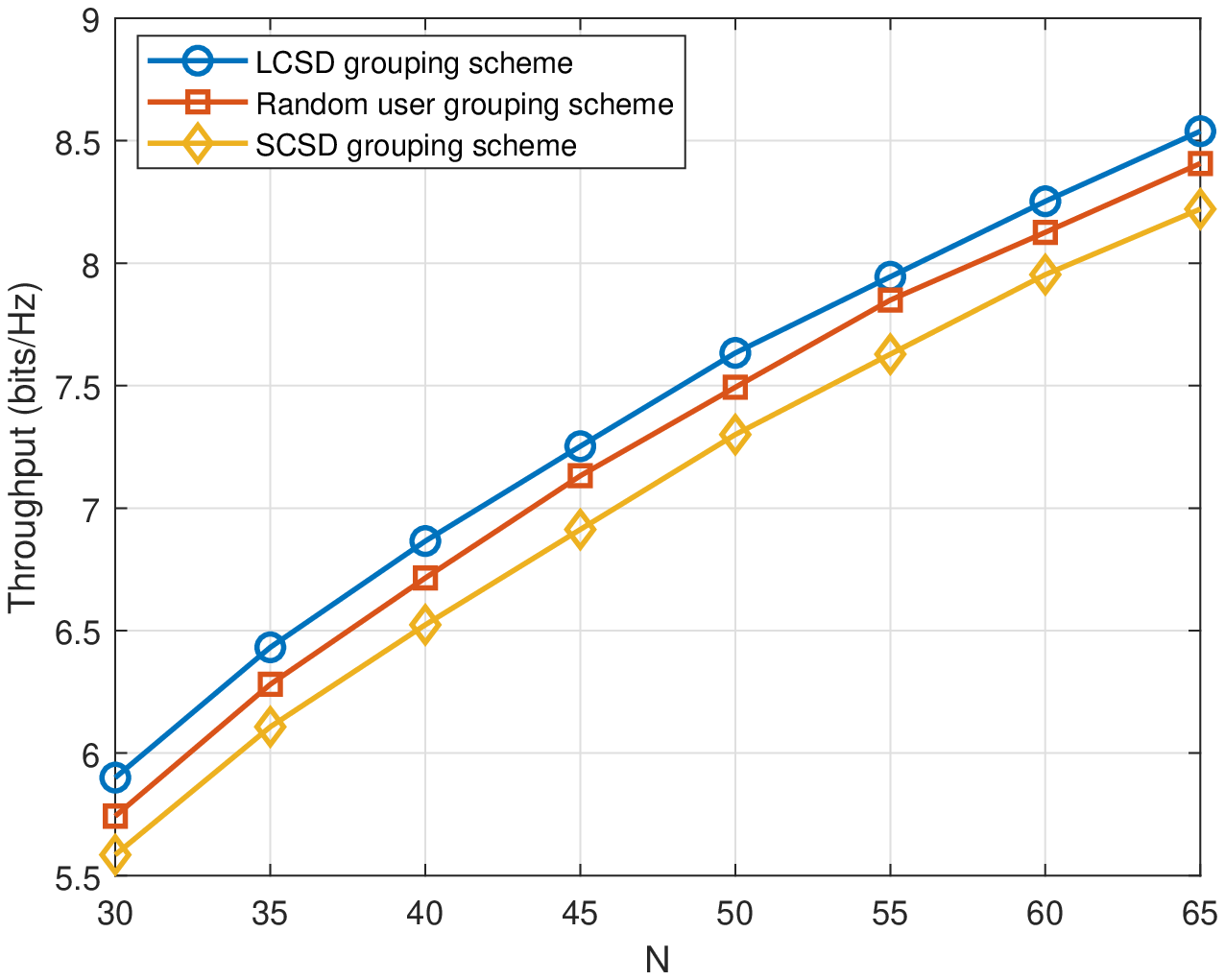} \label{FigThr_Cluster}} 
\hfil
\subfloat[Impact of number of user groups.]{\includegraphics[width=0.23\textwidth]{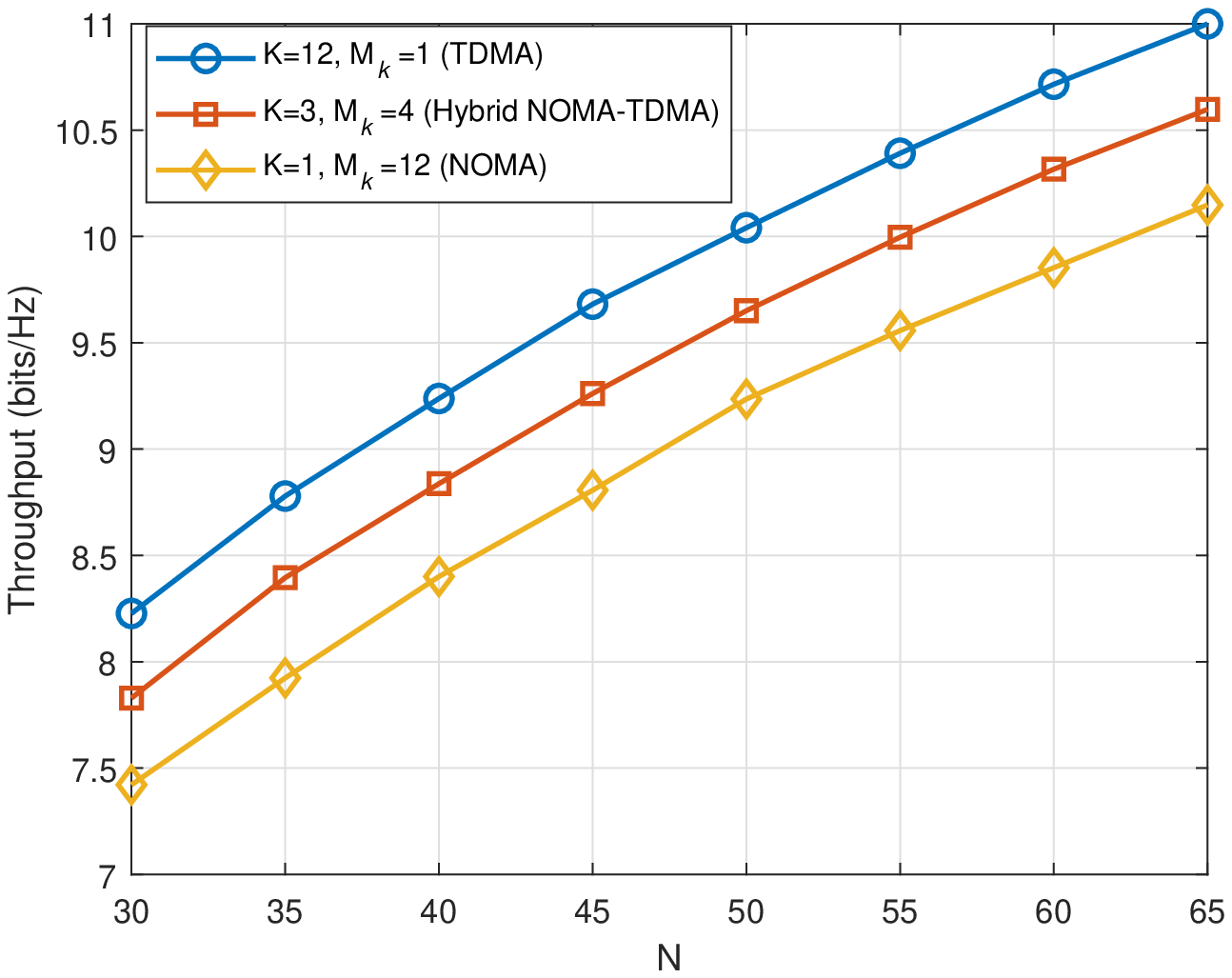} \label{FigThr_Setup}} 
\caption{Simulation results.}  
\label{fig_sim}
\end{figure*}

First, we randomly group all users into $K=3$ clusters with $M_k = 4$ users in each cluster, and show the throughputs of different algorithms versus the IRS's reflecting element number $N$ and the distance between the IRS and the users $d_{k,m}$ when $P_0=40$ \text{dBm} in Figs. 2(a) and 2(b), respectively. $d_{k,m}=5$ m in Fig. 2(a), and $N=64$ in Fig. 2(b). It is observed that the throughput of the proposed algorithm is very close to its upper bound and is significantly higher than that of the other benchmark algorithms. The proposed algorithm outperforms the `Optimized IRS w/ TA (same IRS)' algorithm, because it has more degrees of freedom for optimization than the latter. Besides, the proposed algorithm with continuous IRS phase shifts always outperforms its `Discrete phase' counterpart with $b=1, 2, 3$ control bits, and the performance gap between them decreases as $b$ grows. This is because signal misalignment caused by low-resolution phase shift degrades system performance. Furthermore, the throughput gain of the proposed algorithm over the 'Optimized IRS w/o TA' algorithm and that of the `Random IRS w/ TA' algorithm over the `Random IRS w/o TA' algorithm demonstrate the effectiveness of time allocation in improving throughput. Besides, the throughput gain of the proposed algorithm over the `Random IRS w/ TA' algorithm demonstrates the effectiveness of IRS reflecting beamforming in improving throughput. Thus, the proposed algorithm can fully reap the gains brought by IRS reflect beamforming and time allocation.

Fig. 2(c) shows the throughput of the proposed algorithm under three different user grouping schemes versus $N$ under the same condition with Fig. 2(a) except that $d_{k,m}$ is random in $[5,15]$ m. To describe the user grouping schemes, we rename and order the users as $u_1, \ldots, u_N$ such that $||\mathbf{h}_1|| \geq \ldots \geq ||\mathbf{h}_N||$, where $\mathbf{h}_n$ denotes the channel coefficient from user $n$ to the IRS, $n=1,\ldots,N$. The first user grouping scheme is called large-channel-strength-difference (LCSD) scheme, which lets the user-IRS channel strength differences among the users in the same cluster as large as possible, and the user clusters obtained by this scheme can be expressed as $\{ u_1, u_4, u_7, u_{10} \}$, $\{ u_2, u_5, u_8, u_{11} \}$, and $\{ u_3, u_6, u_9, u_{12} \}$. The second scheme is called small-channel-strength-difference (SCSD) scheme, which lets the user-IRS channel strength differences among the users in the same cluster as small as possible, thus its obtained user clusters can be expressed as $\{ u_1, u_2, u_3, u_{4} \}$, $\{ u_5, u_6, u_7, u_{8} \}$, and $\{ u_9, u_{10}, u_{11}, u_{12} \}$. The third scheme is called random user grouping scheme, which groups the users randomly. It is observed from Fig. 2(c) that the LCSD scheme achieves the highest throughput while the SCSD scheme achieves the lowest. This is because, as compared to the other two schemes, the LCSD scheme achieves the smallest overall channel strength difference among different user clusters and thus can achieve the highest performance gain via IRS reflect beamforming and time allocation optimization.

Next, we show the throughput of the proposed algorithm under three different user clustering setups versus $N$ when $P_0=40$ dBm and $d_{k,m}=5$ m in Fig. 2(d), where all users are randomly grouped into $K=12/3/1$ clusters with $M_k = 1/4/12$ user(s) in each cluster. It is observed that the more clusters that the users are grouped, the higher throughput can be achieved. This is because, with more user clusters, there are more degrees of freedom for both IRS reflect beamforming and time allocation optimization, which results in higher throughput.

\section{Conclusion}
This letter studied an IRS-assisted wireless powered hybrid NOMA and TDMA network, where IRS reflect beamforming and time allocation are jointly designed to maximize the throughput of the network. Based on the BCA, SDR, and SROCR techniques, an efficient algorithm was proposed to solve the considered challenging problem. Simulation results have shown that the proposed algorithm achieves significantly higher throughput than other benchmark algorithms. It is also found that grouping users into more clusters brings more degrees of freedom for the joint IRS beamforming and time allocation and thus can achieve better throughput performance.


\end{document}